\documentclass[11pt,twoside]{article}
\usepackage{asp2014}

\aspSuppressVolSlug
\resetcounters

\begin{document}

\title{ALMA's high-cadence imaging capabilities for solar observations}

\author{Sven Wedemeyer,$^{1,2}$ Asbj{\o}rn Parmer,$^1$
\affil{$^1$Institute of Theoretical Astrophysics, University of Oslo, Norway;  \email{sven.wedemeyer@astro.uio.no}}
\affil{$^2$European ALMA Regional Center, Ondrejov, Czech Republic} 
}

%

\paperauthor{Sven~Wedemeyer}{sven.wedemeyer@astro.uio.no}{http://orcid.org/0000-0002-5006-7540}{University of Oslo}{Institute of Theoretical Astrophysics}{Oslo}{}{0315}{Norway}
\paperauthor{Asbjorn~Parmer}{}{}{University of Oslo}{Institute of Theoretical Astrophysics}{Oslo}{}{0315}{Norway}

\begin{abstract}
The Atacama Large Millimeter/submillimeter Array offers an unprecedented view of our Sun at sub-/millimeter wavelengths. 
The high spatial, temporal, and spectral resolution facilitates the measurement of gas temperatures and magnetic fields in the solar chromosphere with high precision. 
The anticipated results will revolutionize our understanding of the solar atmosphere and may in particular result in major steps towards solving the coronal heating problem. 
Based on state-of-the-art 3D radiation magnetohydrodynamic simulations, we calculate the emergent continuum intensity (and thus brightness temperature maps) in the wavelength range accessed by ALMA and simulate instrumental effects for different array configurations. 
First results show that the local gas temperature can be closely mapped with ALMA and that much of the complex small-scale chromospheric pattern can be resolved. 
\end{abstract}

%
%
\section{Predictions for ALMA observations of the quiet Sun}
%
The presented calculations utilize a model of a quiet Sun region \citep{2012Natur.486..505W}, which is the result of a realistic state-of-the-art 3D radiation magnetohydrodynamic (MHD) simulation with CO5BOLD \citep{2012JCoPh.231..919F}. 
It comprises the upper part of the solar convection zone, the photosphere and the chromosphere. 
The model can reproduce a large number of key observables, e.g., the properties of the granulation in the low photosphere.
The chromosphere above, which is the layer mapped with ALMA, is characterized by an intricate interplay of propagating shock waves and magnetic fields. 
The resulting dynamic pattern features hot shock fronts and cool post-shock regions on spatial scales down to a few 0.''1 and on time-scales of only a few seconds. 
The modelled gas temperatures are in the range between 2000\,K and 8000\,K.

\articlefigure[width=.8\textwidth]{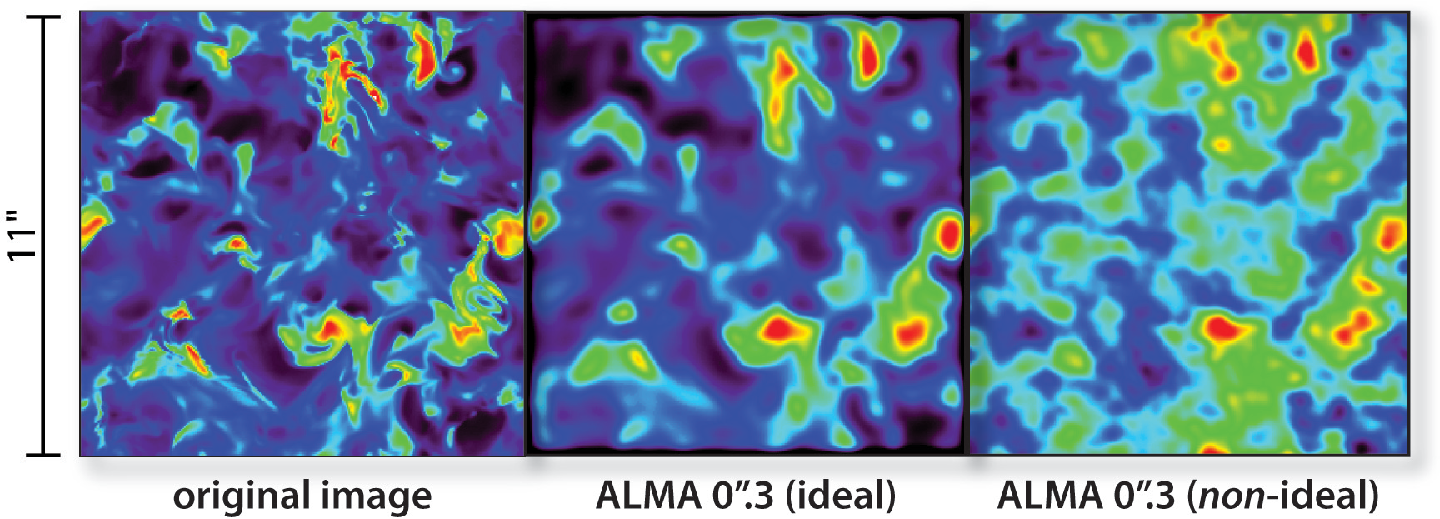}{fig1}{The original brightness temperature map at 1\,mm (left) and the CASA results after applying an ideal (middle) and the full (non-ideal) PSF (right), corresponding to an ALMA configuration for an effective angular resolution of 0.''3.}

The numerical model (available as a time sequence with 1\,s cadence) is used as input for radiative transfer calculations with LINFOR3D\footnote{Steffen, Ludwig, Wedemeyer: \url{http://www.aip.de/~mst/linfor3D_main.html}}, which outputs the emergent continuum intensity in the ALMA wavelength range.
The resulting brightness temperature maps show an intermittent and dynamic pattern very similar to the original gas temperature maps in the model chromosphere \citep[cf.][]{2007A&A...471..977W,loukitchevabifrostmm}. 
The synthetic maps are then used as input for CASA. 
Different array configurations are considered. 
Convolution of the maps with the point spread function (PSF) for an ideal synthesized aperture (``clean beam'') and with the more realistic non-ideal PSF result in artificial observations, i.e. images like they  would be expected from ALMA based on the input model (see Fig.~\ref{fig1}).

%

%
\section{Results and conclusion}
%
The analysis confirms that the local gas temperature can be closely mapped with ALMA and that the effective formation height of the radiation increases with wavelength, which turns ALMA into a tomograph. 
At the shortest wavelength (0.3\,mm) ALMA maps the top of the photosphere, whereas the intensity at the longest wavelengths (up to 8.6\,mm) is formed in the upper chromosphere.
Although the smallest scales ($\sim 0''.1$) may not be resolved consistently at 1\,mm, ALMA will be capable of mapping the thermal structure of the solar chromosphere at a resolution not far from what is possible with the currently best optical solar telescopes. 
In combination with the high spectral and temporal resolution, this turns ALMA into a tool that can perform unprecedented measurements, which may result in fundamental new insights and steps towards solving a wide range of problems in solar physics, incl. the solar coronal heating problem.
See \cite{ssalmon_ssrv15} for a comprehensive discussion of potential solar science with ALMA.
%
%
\acknowledgements The presented results are part of the master thesis of A.~Parmer (2015, Univ. of Oslo, Norway), carried out in connection with the SSALMONetwork\footnote{Please refer to \cite{ssalmon_espm15} and the network webpages at \url{http://ssalmon.uio.no} for more information.}.
\vspace*{-7mm} 


\vspace*{-5mm}
\end{document}